\documentclass[11pt,a4paper]{article}
\usepackage{mathrsfs}
\usepackage{amsmath}
\usepackage{amssymb}
\usepackage{mathdesign}
\usepackage{dsfont}
\usepackage{esint}
\usepackage{youngtab}

\usepackage{amsmath,amssymb,amsfonts,a4wide,graphicx,bm,times,psfrag,wrapfig,sidecap}
\usepackage{cite}
\usepackage[colorlinks=true,linkcolor=black, citecolor=black,
urlcolor=black]{hyperref}
\numberwithin{equation}{section}
\makeatletter \let\old@startsection=\@startsection
\renewcommand{\@startsection}[6]
{\old@startsection{#1}{#2}{#3}{#4}{#5}{#6\mathversion{bold}}}
\makeatother

\def\be{\begin{eqnarray}  }
    \def\ee{\end{eqnarray}}

\def\({\left(} \def\){\right)}
\def\<{\langle}
\def\>{\rangle}
\def\[{\left[}
 \def\]{\right]}

\newcommand\encadremath[1]{\vbox{\hrule\hbox{\vrule\kern8pt
\vbox{\kern8pt \hbox{$\displaystyle #1$}\kern8pt}
\kern8pt\vrule}\hrule}} \def\enca#1{\vbox{\hrule\hbox{
\vrule\kern8pt\vbox{\kern8pt \hbox{$\displaystyle #1$} \kern8pt}
\kern8pt\vrule}\hrule}}

  \usepackage{bm}

\def\ee{\end{eqnarray}}  

\def\({\left(} \def\){\right)}   \def\[{\left[} \def\]{\right]}

\def\Re{{\rm Re}}
\def\Im{{\rm Im}}

\begin{document}

\thispagestyle{empty}

\begin{flushright}

\end{flushright}

\vspace{1cm}
\setcounter{footnote}{0}

\begin{center}

\begin{center}

{\Large\bf Viscous fingering in the presence of weak disorder}

\vspace{7mm}
{Eldad Bettelheim \& Oded Agam} \\[5mm]

{\it Racah Inst.  of Physics, \\Edmund J. Safra Campus,
Hebrew University of Jerusalem,\\ Jerusalem, Israel 91904 \\[5mm]}

\end{center}

\abstract
We consider the problem of viscous fingering in the presence of quenched disorder that is both weak and short-range correlated. The two point correlation function of the harmonic measure is calculated perturbatively, and is used in order to calculate the correction the the box-counting fractal dimension. We show that the disorder increases the fractal dimension, and that its effect decreases logarithmically with the size of the fractal.  
\end{center}

\section{Introduction}

The phenomenon of viscous fingering, realized, e.g., when a fluid (say water) displaces another  more viscous  fluid (say oil) within a constricted geometry \cite{Saffman-Taylor}, such as porous medium or Hele-Shaw cell \cite{Hele-Shaw}, is a central paradigm in non-equilibrium and pattern formation physics. Its complexity arises from the long range interactions and the screening effects of the developing fingers.  The non local effects manifest themselves usually in fractal structures that are still not entirely understood \cite{Book, Mineev}. 

Most of our knowledge about these problems comes from extensive numerical studies which, in particular,  show that anisotropy, surface tension, and quenched disorder, have an important impact on the  geometric properties \cite{Review}. The effect of quenched disorder - the focus of this work - has been studied mainly in two limits: The limit of strong disorder  (long range correlated) where the percolative nature of the dynamics becomes dominant \cite{Strong4,Strong3,Strong2,Strong15,Strong13, Strong1}, and the weak disorder limit which was studied mainly in a channel geometry \cite{Weak1,Weak2,Weak3,Weak4, Weak5}.  The focus in the latter case was on the roughness and growth exponents of the developing interface, in the presence of capillary effects which introduce additional length scales to the problem.  

In this work we consider the effect of weak disorder on viscous fingering in two dimensional space such as the Saffman-Taylor problem \cite{Saffman-Taylor} in a Hele-Shaw cell \cite{Hele-Shaw} , neglecting surface tension. Disorder in this system can be introduced, e.g., by varying the gap width, $h$, between the two plates of the cell. Assuming the  variation  of the size of the gap between the plates to be smooth, $|\nabla h| \ll 1$, Darcy's law holds locally, i.e.
\begin{equation} 
 {\bm v}= -\frac{h^2}{12 \mu } {\bm \nabla} P, \label{velocity}
\end{equation} 
where ${\bm v}$ is the two dimensional velocity vector averaged over the cell gap, $h=h({\bm r})$ is the local gap { height},  $\mu$ is the viscosity, and $P$ is the pressure. We define the dimensionless gap { height}:
\begin{align}
\eta(\bm{r})=\frac{h(\bm r)}{h_0},
\end{align} 
where $h_0$ is the average gap height. Assuming incompressible flow, we have
\begin{equation} 
\bm \nabla \cdot \left(\eta {\bm v} \right) =Q\delta({\bm r}-{\bm r}_s),
\end{equation}
where the multiplication of the velocity by $\eta$ accounts for conservation of fluid volume, and  the right hand side represents  a source term located at ${\bm r}_s$ . This source  accounts for the expansion of the Saffman-Taylor bubble at a rate $Q$ (with units of area over time). From the above equations it follows that
\begin{equation} 
\bm \nabla \cdot \left( D \bm \nabla P\right) = Q\delta({\bm r}-{\bm r}_s), \label{basicEq}
\end{equation}
with a space-dependent diffusion constant
\begin{equation} 
D({\bm r})= D_0\eta^3(\bm r)\simeq D_0\left(1+ 3\delta \eta(\bm r) \right).
\end{equation}
Here $D_0=\frac{h_0^2}{12 \mu}$ is the average value of the diffusion constant, while  $\delta \eta({\bm r})=\eta(\bm r)-1     $ denotes the random variations in the gap size, assumed to be small $|\delta \eta(\bm r)|\ll 1$. 

In what follows we shall extend   the traditional Saffman-Taylor problem in a Hele-Shaw cell by introducing a weak and short range correlated disorder in the gap width of the cell. We assume weak disorder with zero mean $\<\delta \eta({\bm r})\>=0$ and a short-range correlation function:
\begin{align}
\quad\<\delta \eta({\bm r})\delta \eta({\bm r'}) \> =  g(\bm r- \bm r'),
\end{align}
(For our perturbative treatment higher correlations of the disorder are irrelevant). We further assume that for large enough distances, $|\bm r-\bm r'|\gg\sigma$, the two point correlation function, $g(\bm r-\bm r'), $ may be replaced by a delta function $g(\bm r-\bm r')\to \ell^2\delta(\bm r-\bm r')$, while for short distances,$|\bm r-\bm r'|\ll\sigma $, the correlation function approaches a constant $g(\bm r-\bm r')\to \ell^2/\sigma^2$.  Here $\ell \sim \epsilon \sigma$ denotes the disorder length scale, which depends  both on  the relative variation of the gap   $\epsilon \sim |\delta\eta| \ll 1$ as well as the disorder correlation length, $\sigma$. The limit of smooth gap variations implies  $h_{0}|\delta \eta| /\sigma \ll 1$. 

A central quantity characterizing Laplacian growth is the harmonic measure,  associated with the probability of growth in a unit time along some point on the boundary of the Saffman-Taylor  bubble. It is, essentially, the component of the velocity which is normal to the boundary:
\begin{equation}
v_n({\bm r})= -D_0\eta^2(\bm r) \frac{\partial P({\bm r})}{\partial n} \label{v_n} 
\end{equation}
where ${\bm r}$ is a point on the boundary of the bubble, while $\partial_n$ represents the normal derivative to the boundary, at the corresponding point. A basic quantity which characterizes the effect of  disorder on the dynamics of the growing Saffmat-Taylor bubble is the fluctuation  of the normal velocity  at the boundary  bubble:\begin{equation}
\delta v_{n}({\bm r})=  v_{n}({\bm r})- \<v_{n}({\bm r})\>.
\end{equation} 

The  goal of this work is characterize the statistics of  $\delta v_{n}({\bm r})$ and to use it in order to quantify the effect of disorder on the fractal dimension of the developing patterns. For this purpose it will be sufficient to consider  the case  of a bubble in the form of a wedge with an opening angle $\phi$, where the limit $\phi=0$ corresponds to a sharp tip, while $\phi= 2\pi$ corresponds to a sharp fjord. 

To explain the motivation for this choice, let us recall the characterization of a fractal pattern using the multi-fractal spectrum function $f(\alpha)$ \cite{efofalfa}. Suppose the boundary of a fractal pattern is covered by boxes of size $\xi$. Then the rate of growth probability within the $i$-th box scales as $p_i \propto\xi^\alpha$, and the frequency of observing a particular value of $\alpha $ within a range $d \alpha$ in all the boxes is proportional to $\xi^{-f(\alpha)} d\alpha$. 

Now, following Ref.~\cite{Wedge}, we  argue that
each value of $\alpha$ corresponds to a wedge of opening angle 
 \begin{align}
\phi =2\pi- \frac{\pi}{\alpha}, \label{phi}
\end{align}
where $\alpha$ changes between $1/2$ ( for a sharp tip) and $\infty$ (for a sharp fjord). 

Consider the harmonic measure of a wedge with opening angle $\phi$  in a non-disordered system. The pressure, $P({\bm r}),$ satisfies Laplace equation except at the sources locations, and therefore may be expressed by the imaginary part of an analytic conformal mapping, $w(z)$, from  the exterior of  the bubble domain ($z$-plain) to, say, the upper half plane ($w$-domain):
\begin{align} 
P({\bm r})=-\frac{Q}{2\pi D_0} \Im [w(z)].
\end{align}
(We shall use the convention that the mathematical plane variable, $w$\ , is dimensionless and the physical plane variable, $z$, has the dimensions of length.)

The normal derivative, at the boundary of the bubble, can be obtained from  
\begin{align}
&\frac{\partial}{\partial n}=i\left|\frac{\partial w}{\partial z} \right|\left(
\frac{\partial}{\partial w}-\frac{\partial}{\partial \bar{w}}\right) \label{Normal-D}
\end{align} 
where overbar denotes complex conjugation. Thus the harmonic measure in the absence of disorder is
\begin{equation}
v_n^{(0)}({\bm r})= -D_0 \frac{\partial P({\bm r})}{\partial n}=\frac{Q}{2\pi }\left|\frac{\partial w(z)}{\partial z} \right|. \label{v_n0}
\end{equation}

The conformal mapping from the upper half plane ($w$-plane) to a wedge ($z$-plane) is given by
\begin{align}
z(w)=Lw^{ \frac{ 2\pi-\phi}{\pi}}=Lw^{ \frac{ 1}{\alpha}}. \label{Conf_wedge}
\end{align}
where  $L$ is some constant with dimensions of length. The wedge boundary in the target plane, $z=x+iy$, corresponds to the real axis in $w$ plane. In particular, the positive real axis in $w$-plane is mapped to the positive real axis in $z$-plane, while the negative real axis of $w$-plane is mapped to a line forming an angle $\phi$ with the $x$-axis, as shown in Fig.~1.
\begin{figure}[t]
    ~~~~~~~~~~~~~~~~~~~~~~~~\includegraphics[width=0.35\columnwidth]{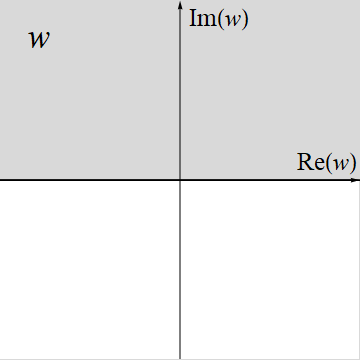}~~~~~
    \includegraphics[width=0.35\columnwidth]{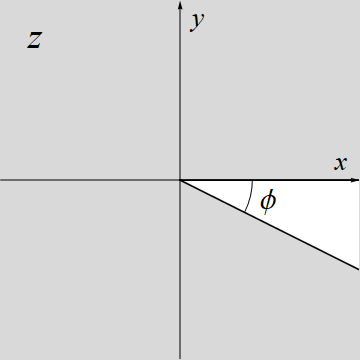}
\caption{The conformal mapping from the upper half plane (left panel) to the exterior (gray area) of a wedge of angle $\phi$ (right panel). }
 \label{WedgeConformalMapFig}
\end{figure}
From Eqs.~(\ref{v_n0})\ and (\ref{Conf_wedge}) it follows that
\begin{equation}
v^{(0)}_n = \frac{\alpha Q}{2\pi} \frac{|z|^{\alpha-1}}{L^\alpha}. \label{HamonicMeasure}
\end{equation} 
Thus  probability rate for a growth of the wedge-bubble    in a box of  size $\xi$, centered at the tip of the wedge, is  
\begin{equation}
p  =\int_{-\xi/2}^{\xi/2} dz v^{(0)}_n = \frac{Q}{\pi}\left(\frac{\xi}{2L}\right)^\alpha, \label{p}
\end{equation}
where the integration is along the boundary of the wedge near its tip (where the  origin of the coordinate system is set). Recall that  $f(\alpha)$\ is defined through  $p\sim\xi^{-f(\alpha)}$, thus computing $p$ will allows one to find $f(\alpha)$.

In the presence of weak disorder, the growth law of a wedge changes, and $p$ cannot be associated with a particular value of $\alpha$. Instead it assumes a narrow distribution of $\alpha$. Yet, for  perturbative description it is sufficient to account for the average value of $\alpha$, i.e. the effective value $\alpha_{eff}$ associated with a wedge of a given angle,
\begin{equation}
 \alpha_{eff}=\alpha- \Delta\alpha(\alpha, \xi). \label{alpha_eff}
\end{equation}
The latter can be deduced from an expectation value involving the harmonic moments associated with a wedge of opening angle (\ref{phi}), therefore
\begin{align}
 &\Delta\alpha(\alpha, \xi)=\alpha-\lim_{\xi/L \to 0  } \frac{ 1  }{\log (\xi/L)} \left\langle \log\int_{-\xi/2}^{\xi/2}  v_n dl\right\rangle  \nonumber=\\&=-\lim_{\xi/L \to 0  } \frac{ 1  }{\log (\xi/L)} \left\langle \log\left(1+\frac{\int_{-\xi/2}^{\xi/2}  \delta v_n dl}{\int_{-\xi/2}^{\xi/2}   v^{(0)}_n dl}\right)\right\rangle    \label{integral},
\end{align}     
where we have used Eq. (\ref{p}) to obtain the last equality, and $\delta v_n= v_n-v_n^{(0)}$ with $v_n^{(0)}$ defined in (\ref{HamonicMeasure}). 

Consider now the fractal dimension  (the box counting dimension) of a developed Saffman-Taylor bubble. It may be calculate by enumerating the number of boxes $N(\xi)$ of size $\xi$ that cover the bubble of linear size $L_{\max}$. Namely,
\begin{equation}
d_0  = \lim_{\xi/L_{\max} \to 0} \frac{\log(N(\xi))}{\log (L_{\max}/\xi)}
\end{equation}
Viewing the fractal bubble as composed of wedges of different opening angles, the box number $N(\xi)$ can be expressed as an integral over $\alpha$,
\begin{equation}
N^{}(\xi)  = \int d\alpha \rho(\alpha)\left(\frac{L_{\max}}{\xi}\right)^{f(\alpha)},
\end{equation}
where $\rho(\alpha)$ is some smooth function of $\alpha$. In the limit $L_{\max} \gg \xi$, this integral is governed by its saddle point $f'(\alpha)=0 $, therefore $N(\xi) \sim (L_{\max}/\xi)^{f_{\max}}$, where $f_{\max}$ is the maximal value of $f(\alpha)$.  Thus the fractal dimension is this maximal value, $d_0= f_{\max}$.

Formula (\ref{integral}) can now be used  in order to calculate, perturbatively, the correction of the disorder to the fractal dimension, $d_0$.  From the relation $(L_{\max}/\xi)^{f(\alpha)} d\alpha= (L_{\max}/\xi)^{\tilde{f}(\alpha_{eff})}d\alpha_{eff}$, and definition (\ref{alpha_eff}) we obtain that the distribution function which takes into account  the effect of the disorder is
\begin{equation} 
\tilde{f}(\alpha_{eff})= f(\alpha)-\frac{1}{\log (\xi/L_{\max})}  \frac{ \partial \Delta \alpha}{\partial \alpha}
\end{equation}
Let  $\alpha_{eff}^*$ be the point where $\tilde{f}(\alpha_{eff})$ reach its maximal value, and assume it  corresponds to $\alpha^*-\Delta\alpha(\alpha^*,\xi)$, where $\alpha^*$ is the maximal point of $f(\alpha)$ (in the absence of disorder). Since $f(\alpha^*)=d_0$ and $f'(\alpha^*)=0$, we obtain that, to leading order in the strength of the disorder, the effective fractal dimension is:
\begin{equation}
d_0^{(eff)}(\xi)= d_0- \frac{1}{\log (\xi/L_{\max})}\frac{ \partial \Delta \alpha(\alpha^*,\xi)}{\partial \alpha^*}. \label{dd}
\end{equation}
  
 \section{Perturbation theory}
In this section we  derive the perturbative formula for $\Delta\alpha(\alpha,\xi)$. Let us define the Green function associated with Eq.~(\ref{basicEq}):
\begin{align}
-\left[\bm \nabla^2 + 3\bm \nabla( \delta \eta({\bm r})\cdot \bm \nabla)\right] G({\bm r},{\bm r}')=  \delta({\bm r}-{\bm  r}')
\end{align}
Then, assuming the source to be at infinity, ${\bm r}' \to \infty$, the solution for the pressure is
\begin{align}
P({\bm r})= \frac{Q}{D_0} G({\bm r},\infty), \label{P_r}
\end{align}
and therefore
\begin{align}
v_n({\bm r})= -Q\eta^2(\bm r) \frac{\partial }{\partial n}G({\bm r},\infty). \label{P_r}
\end{align}
Substituting this formula in Eq.~ (\ref{integral}), we express $\Delta \alpha(\alpha,\xi)$ making use of   Green functions: 
\begin{align}   
&\Delta\alpha(\alpha, \xi)\simeq\lim_{\frac{\xi}{L}\to0}  \frac{Q^2}{2p^{2}\log(\xi/L)} \left\<\left[\delta\int_{-\xi/2}^{\xi/2} \eta ^{2}({\bm r}) \frac{\partial G({\bm r},\infty)}{\partial n}dl\right]^2\right\>,\label{DeltaAlphaExpression}
 \end{align}
where $\delta$ denotes the fluctuating part of the following expression. To derive this formula we expand the logarithm in  (\ref{integral}) to second order and use  (\ref{P_r}) and (\ref{p}).

\begin{figure}[t]
~~~~~~~~~~~~~~~~~~~~~~~\includegraphics[width=0.75\columnwidth]{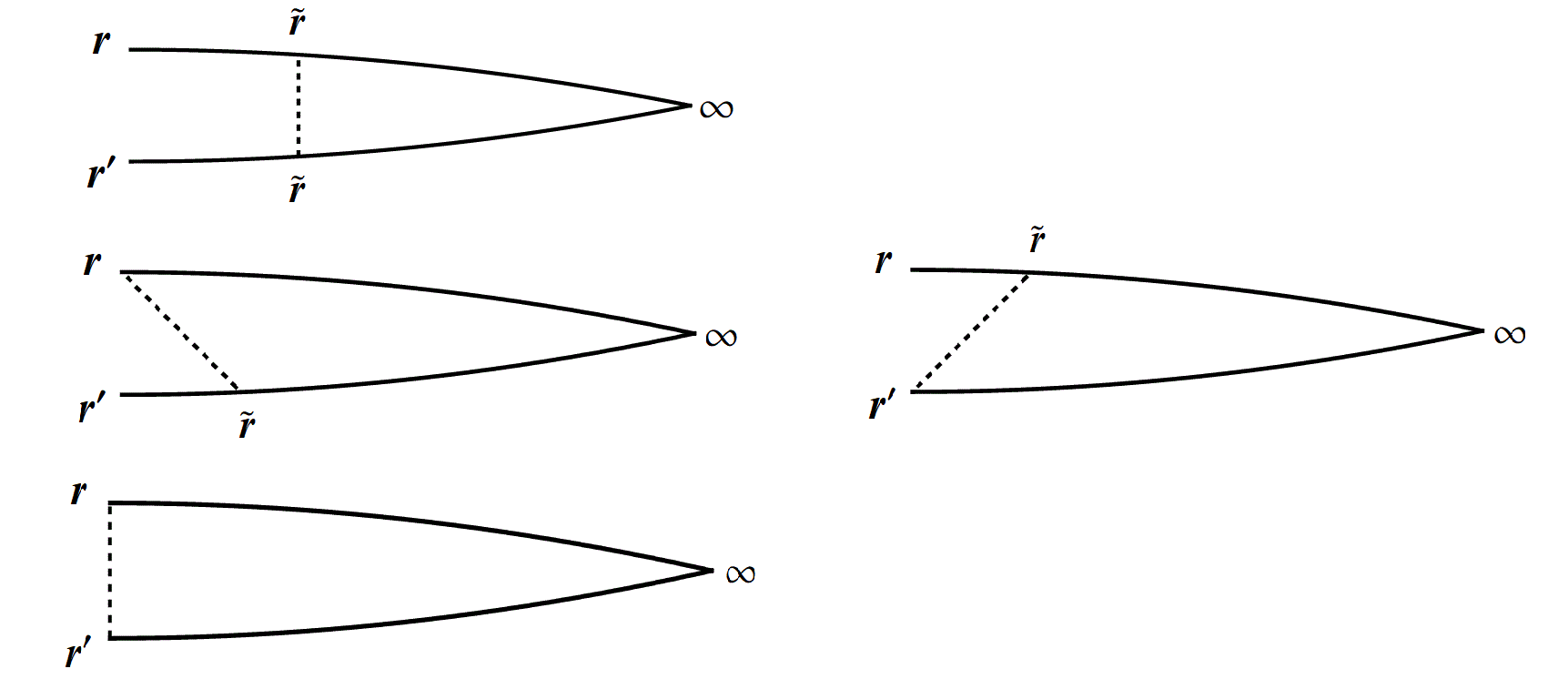}
\caption{The leading order diagrams contributing to $\Delta\alpha(\alpha, \xi)$. }
 \label{diagramsfig}
\end{figure}

To construct the perturbative expansion of the above expression we expand the Green function in a power series $G=\sum_n G_{n}$,
where   the $n$-th term is proportional to the $n$-th power of the perturbation:
\begin{align}
V =3\bm \nabla( \delta \eta(\bm r)\cdot \bm \nabla), 
\end{align}
while the zeroth  order Green function satisfies the equation
\begin{align}
 \quad -\bm \nabla^2G_{0}= \delta ({\bm r} - {\bm r}'),
\end{align}
Thus
\begin{align}
 G_n =G_0 (V G_0)^n.\label{ExpansionG}
\end{align}
The leading order perturbative contribution to $\Delta \alpha (\alpha,\xi)$ comes from a second order expansion in $\delta \eta$. In writing out all these contributions to Eq. (\ref{DeltaAlphaExpression}), one encounters two sources of fluctuations, one is associated with the expansion of the Green function according to (\ref{ExpansionG})  while the other comes from the term  $\eta^2({\bm r})$ appearing in (\ref{DeltaAlphaExpression}).  The different contributions may be enumerated by employing a diagrammatic scheme whereby  solid lines represent the Green function $G_0$ (in
the absence of disorder), while dashed lines represent the disorder correlator $\<\delta \eta(\bm r)\delta \eta(\bm r')\>=g(\bm r- \bm r')$. These diagrams are displayed in Fig. \ref{diagramsfig}. In particular, the contributions to the correlation function of the harmonic measure,
\begin{equation}
C(\bm r, \bm r') = \left\<  \delta \left[ \eta^2(\bm r) \frac{\partial G({\bm r},\infty)}{\partial n}\right] \delta \left[\eta^2(\bm r') \frac{\partial G({\bm r'},\infty)}{\partial n}\right]\right\>,
\end{equation}
associated with diagrams (a,b,c), which we denote as $C^{(a,b,c)}({\bm r},{\bm r}')$ respectively, are: 
\begin{subequations}
\begin{equation}
C^{(a)}({\bm r},{\bm r}')= 9 \ell^2 \frac{\partial^2}{\partial n \partial n'}\int d^2\tilde{r}  \left(\frac{\partial G_{0}({\bm r}, \tilde{\bm r}) }{\partial \tilde{\bm r}}\cdot\frac{\partial G_{0}( \tilde{\bm r}, \infty) }{\partial \tilde{\bm r}}\right) \left(\frac{\partial G_{0}({\bm r}', \tilde{\bm r}) }{\partial \tilde{\bm r}}\cdot\frac{\partial G_{0}( \tilde{\bm r}, \infty) }{\partial \tilde{\bm r}}\right), \label{nl}
\end{equation}

\begin{equation}
C^{(b)}({\bm r},{\bm r}')= -6\frac{\partial G_{0}({\bm r}, \infty)}{\partial n}\int d^2\tilde{r}\frac{                \partial}{\partial n'}\left(\frac{\partial G_{0}({\bm r}',\tilde {\bm r}) }{\partial {\tilde{\bm r}}}\cdot\frac{\partial G_{0}(\tilde  {\bm r}, \infty) }{\partial {\tilde {\bm r}}}\right)g (\tilde{\bm r}-{\bm r})+ \left({\bm r} \leftrightarrow {\bm r'}\right), \label{b}
\end{equation}

\begin{equation}
C^{(c)}({\bm r},{\bm r}')= 4 \ell^2 \left(\frac{\partial G_{0}({\bm r}, \infty)}{\partial n}\right)^2 g({\bm r}-{\bm r}'). \label{C-loc}
\end{equation}
\end{subequations}
For future purpose, in the above formulae for $C^{(b,c)}({\bm r},{\bm r}')$ we use the exact form of the disorder correlation function instead of its approximation using delta function.      

\section{Complex notations}
Let us define the complex angle $\theta(\zeta)=-i\log \zeta$, where $\zeta=x+iy$ is {\ the} coordinate in the complex plane, then the Green function of the diffusion operator that satisfies Dirichlet boundary conditions on the circle $|\zeta|=1$ is
\begin{align} 
G_0(\theta, \theta')=- \frac{1}{2\pi}\log\left|\frac{e^{ i\theta}-e^{i\theta'}}{1- e^{i(\theta- \bar{\theta}')}}\right|,
\end{align}   
where we use the notation $\theta=\theta(\zeta)$, $\theta'= \theta(\zeta'),$ and an overbar denotes, as usual, complex conjugation.  When the source is taken to infinity, $|\zeta'| \to \infty$, we have 
\begin{align} 
G_0(\theta, \infty)=-\frac{1}{2\pi} \Im \theta=\frac{1}{2\pi} \log|\zeta|, \label{G0_infinity}
\end{align}
while if both $\theta$ and $\theta'$ are small compared to 1,  the above green function reduces to the well known green function in the upper half $\theta$-plane:
\begin{align} 
G_0(\theta, \theta') \simeq-  \frac{1}{2\pi}\log\left|\frac{\theta-\theta'}{\theta- \bar{\theta}'}\right|  \label{G_local},
\end{align} 
where $\theta'=- i \log (\zeta')$ is the position of the source. Choosing {\ a} local coordinate system $(x',y')$ such that $\zeta'=1+i(x'+iy')$, and assuming $|x'+iy'| \ll 1$, we have $\theta'\simeq x'+iy'$.  Thus to obtain the\ Green function corresponding to a general geometry described by the conformal mapping from exterior domain of the bubble to the upper half plane $w(z)$ one has to substitute $\theta=w(z)$.

The above description holds when the distances of the observation point $z$ and the source $z'$ from the tip of the wedge are much smaller than some cutoff scale, see Eq.~(\ref{Conf_wedge}). More accurately, the Green function is given by $G_0[\vartheta(z),{\vartheta(z')}]$ where:
\begin{equation}
\vartheta(z)= -i \log [1+iw(z)]. \label{vartheta}
\end{equation}
Thus
in particular\begin{align} 
\frac{\partial G_0(\tilde{\vartheta},\infty)}{\partial\tilde{z}}=\frac{i}{4\pi}\frac{\partial\tilde{ \vartheta}}{\partial\tilde{ z}} ;~~~~\frac{\partial G_0(\vartheta,\tilde{\vartheta})}{\partial \bar{\tilde{z}}}=\frac{-i}{4\pi}\frac{\partial\bar{\tilde{ \vartheta}}}{\partial\bar{\tilde{ z}}} \left[\frac{1}{e^{i(\bar{\tilde{\vartheta}}- \bar{\vartheta})}-1}- \frac{1}{e^{i(\bar{\tilde{\vartheta}}- \vartheta)}-1}\right]. 
\end{align}
From now one we assume that the absolute values of all complex angles are  much smaller than one. Then form the above results we obtain:
\begin{align}
 \frac{\partial G_0({\bm r}, \tilde{\bm r})}{\partial \tilde{\bm r}} \cdot\frac{\partial G_0(\tilde{\bm r},\infty)}{\partial \tilde{\bm r}} \simeq-\frac{1}{(2\pi)^2} \Im \left[\left|\frac{\partial\tilde{ \vartheta}}{\partial\tilde{ z}} \right|^2 \left( \frac{1}{\bar{\vartheta}- \tilde{\vartheta}}-\frac{1}{\vartheta- \tilde{\vartheta}}\right)\right]
\end{align}
and taking the normal derivative of this expression by
\begin{align}
&\frac{\partial}{\partial n}=i\left|\frac{\partial \vartheta}{\partial z} \right|\left(
\frac{\partial}{\partial \vartheta}-\frac{\partial}{\partial \bar{\vartheta}}\right)
\end{align}
we obtain
\begin{align}
\frac{\partial}{\partial n}\left[\frac{\partial G_0({\bm r}, \tilde{\bm r})}{\partial \tilde{\bm r}} \cdot\frac{\partial G_0(\tilde{\bm r},\infty)}{\partial \tilde{\bm r}} \right]
\simeq -\frac{1}{(2\pi)^2}\left|\frac{\partial \vartheta}{\partial z} \right|\left|\frac{\partial\tilde{ \vartheta}}{\partial\tilde{ z}} \right|^2 \Re\left[\frac{1}{(\vartheta- \bar{\tilde{\vartheta}})^2}+\frac{1}{(\vartheta- \tilde{\vartheta})^2}\right].\label{GGdn}
\end{align}
The following formula is also useful:
\begin{align}
\frac{\partial G_0({\bm r},\infty)}{\partial n} =-\frac{1}{2\pi}\left| \frac{\partial \vartheta}{\partial z} \right|.
\end{align}

Eq. (\ref{GGdn}) describes a current source at $\tilde{\theta}$ and an additional image source at $\bar{\tilde{\theta}}$ due to Dirichlet boundary conditions of the pressure on the bubble boundary. Note also that the boundary of the wedge corresponds to the real axis in $w$-plain, therefore observation points on the boundary of the bubble {\ are associated with}  real values of  $\vartheta$, where the expression in the square parenthesis of (\ref{GGdn}) becomes purely real. Substituting the above result  in Eq.~(\ref{nl}), using relation (\ref{HamonicMeasure}), and changing variables  from $\tilde{z}$ to $\tilde{\vartheta}$ we obtain
\begin{align}
 C^{(a)}(z, z')\simeq\frac{9\ell^2}{(2\pi)^4}  \left|\frac{\partial \vartheta}{\partial z} \right|\left|\frac{\partial \vartheta'}{\partial z'} \right| \int d^2\tilde{\vartheta} \left|\frac{\partial\tilde{ \vartheta}}{\partial\tilde{ z}} \right|^2 \left[\frac{1}{(\vartheta- \bar{\tilde{\vartheta}})^2}+\frac{1}{(\vartheta- \tilde{\vartheta})^2}\right]\left[\frac{1}{(\vartheta'- \bar{\tilde{\vartheta}})^2}+\frac{1}{(\vartheta'- \tilde{\vartheta})^2}\right],
\end{align}
where the integration is over the whole complex $\tilde{\vartheta}$ plane. 

Dirichlet boundary conditions of the pressure (i.e. the Green function) manifest themselves as image sources which behave differently in fjords and sharp tips. To illustrate this behavior, in Fig.~3 we depict contour plots of the local behavior of the Green function $G_0^{}[\vartheta(z), \vartheta(z')]$ in $z$-plane for fjord (left panel) and for a tip (right panel), for the case of wedge geometry, $\vartheta(z)= \left(\frac{z}{L}\right)^{\alpha}$ , with $\alpha=4/7$ and $\alpha =4$. In the case of a tip ($\alpha =4/7$), the cut of the function $z^\alpha$ is set to be along the tilted wedge boundary. The image sources in this case extend to the next Riemann sheet.

The contribution to the hamonic measure correlation function due to the  b-type diagram (\ref{b}) can be written in the form 
 \begin{equation}
C^{(b)}(z,z')=-\frac{12\ell^2}{(2\pi)^3}  \left|\frac{\partial \vartheta}{\partial z} \right|\left|\frac{\partial \vartheta'}{\partial z'} \right|  \int d^2\tilde{\sigma}    \frac{(\vartheta'- \vartheta-\tilde{\sigma}')^2- \tilde{\sigma}''^2}{\left[(\vartheta'- \vartheta-\tilde{\sigma}')^2+ \tilde{\sigma}''^2\right]^2}g[\vartheta^{-1}(\tilde{\sigma})] +\left(\vartheta \leftrightarrow \vartheta'\right)
\end{equation}
where $\tilde{\sigma}=\tilde{\sigma}'+i\tilde{\sigma}''$ is a complex coordinate.

\begin{figure}[t]
~~~~~~~~~~\includegraphics[width=0.4\columnwidth]{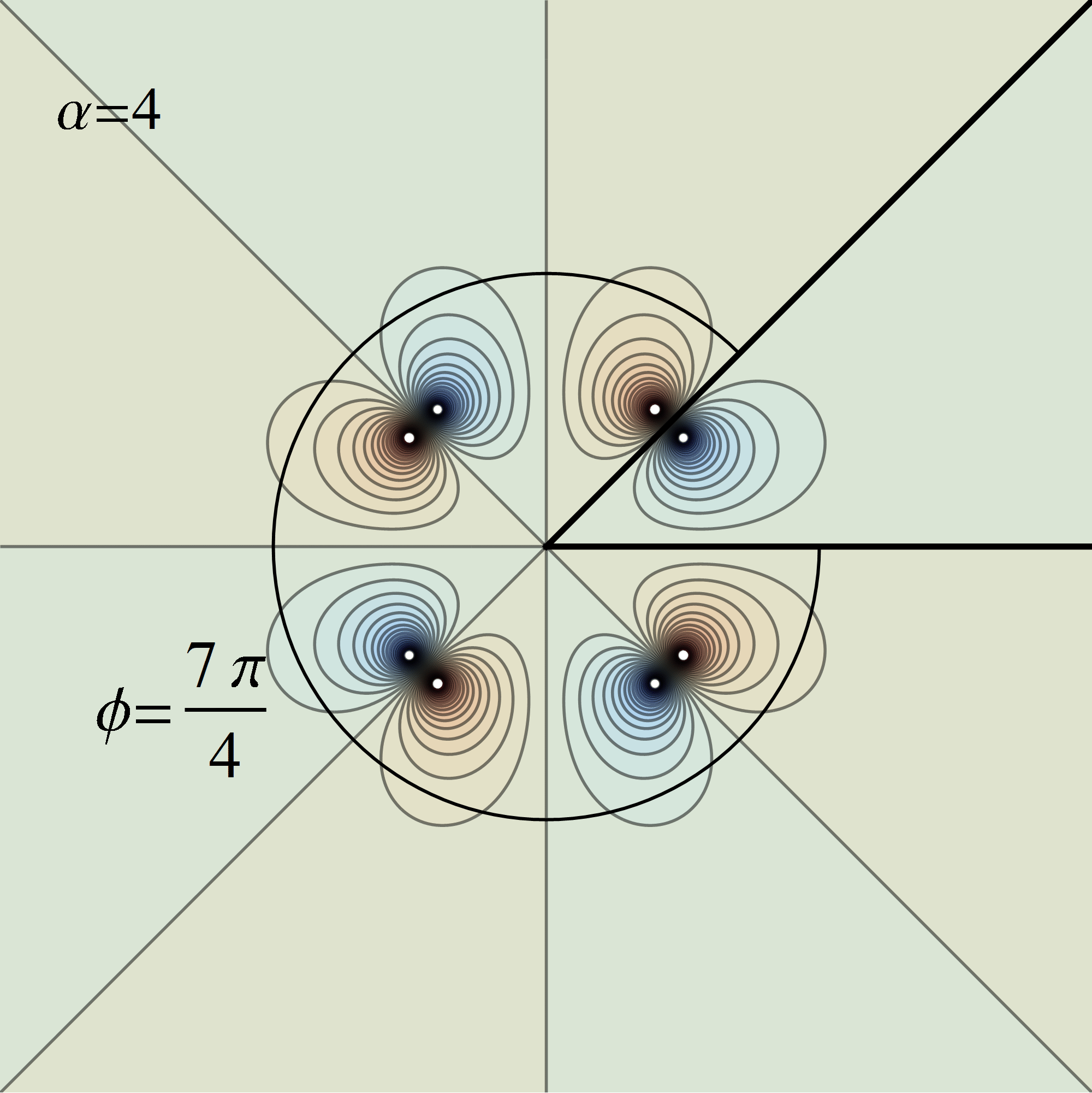}
~~~~~~~~\includegraphics[width=0.4\columnwidth]{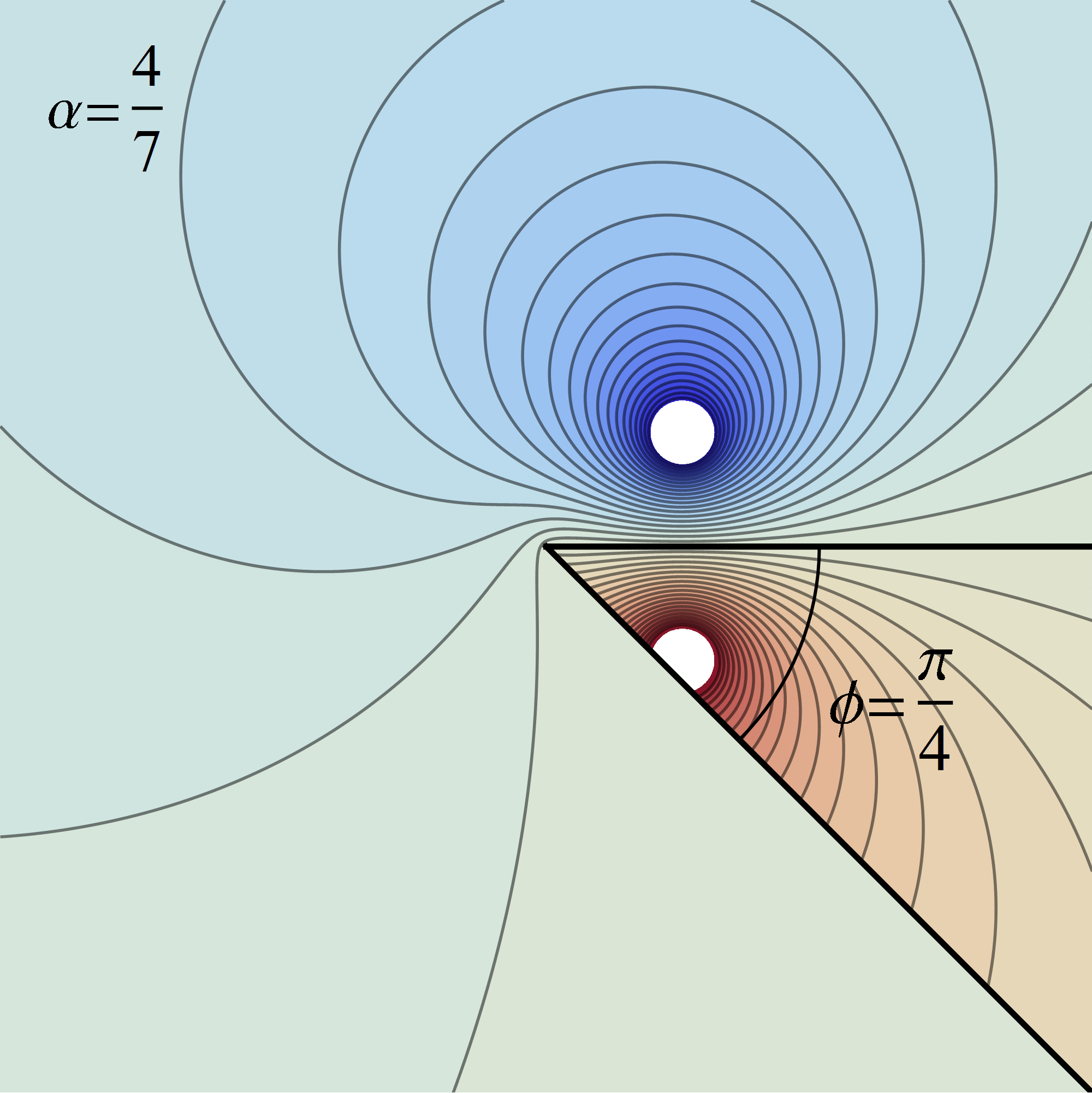}
\caption{A contour plot decribing the local behavior of the Green function $G_0^{}[ \vartheta(z), \vartheta(z')]$ for a fjord (left panel) and for a tip (right panel). The source  in both cases is of the same strength and the same location.}
 \label{GreensFunctionfig}
\end{figure}

Finally, the contribution from the c-diagram (\ref{C-loc}), can be deduced from Eqs. (\ref{G0_infinity}), (\ref{Normal-D}) and  (\ref{v_n0}):
\begin{equation}
C^{(c)}(z,z')=  \frac{1}{\pi^2}  \left|\frac{\partial \vartheta}{\partial z} \right|\left|\frac{\partial \vartheta'}{\partial z'} \right|  g(z-z'), \label{C-locz}
\end{equation}
where $z$ and $z'$ are on the boundary of the bubble.
\section{The fractal dimension}
In what follows we use the above formulae in order to calculate the correction to the fractal dimension. The calculation will be performed for a wedge geometry where $\vartheta(z)= \left(\frac{z}{L}\right)^{\alpha}$, and therefore  $|\frac{\partial \tilde{\vartheta}}{\partial \tilde{z}}|=\frac{\alpha}{L}|\tilde{\vartheta}|^{1-\frac{1}{\alpha}}$. To this end one should, first, calculate the integral over $C^{(a,b,c)}$ along the wedge boundary:
\begin{equation}
I^{(a,b,c)}(\alpha, \xi)= \int_{-\xi/2}^{\xi/2} dz \int_{-\xi/2}^{\xi/2} dz' C^{(a,b,c)}(z,z').
\end{equation}To perform these integrals in this case it will be convenient to change variables from $z$ to $\vartheta$, and from $z'$  to $\vartheta'$.  Consider, first, the contribution from a-diagram: 

\begin{align}
 I^{(a)}(\alpha, \xi)=\frac{ 9 \ell^2 \alpha^2}{(2\pi)^4L^2}  \int_0^\infty d\rho  \rho^{3-\frac{2}{\alpha}} \int_0^{2\pi} d\varphi\left[\frac{4\beta[\beta^2-\rho^2 \cos(2\varphi)]}{\beta^4+\rho^4-2\beta^2 \rho^2 \cos(2\varphi)}\right]^2
\end{align}
where $\beta= (\xi/2L)^\alpha$, and we have used polar coordinates $\tilde{\vartheta}=\rho e^{i\varphi}$ for the integral over $\tilde{\vartheta}$. Integrating over the angle, $\varphi$, gives:
\begin{align}
 I^{(a)}(\alpha, \xi)=\frac{ 9 \ell^2 \alpha^2}{2\pi^3L^2}  \int_0^\infty d\rho \frac{\rho^{3-\frac{2}{\alpha}}}{\beta^2} \left[1+\frac{3\beta^4-\rho^4}{(\beta^2+\rho^2)|\beta^2-\rho^2|}\right]
\end{align}
This integral converges for all $\alpha>1/2$  at $\rho \to \infty$. However, it diverges logarithmically at $\rho=\beta$. This divergence comes from the case where the source of fluctuations approaches the boundary of the bubble, therefore its should be cut off at a distance of order of the disorder correlation length, i.e. at $\rho_{\min}=(\sigma/L)^\alpha$. Expanding the integrand near the divergence point and performing the integral we obtain:
\begin{subequations}
\label{Cintegrals}
\begin{align}
 I^{(a)}(\alpha, \xi)\simeq\frac{ 9 \ell^2 \alpha^3}{2\pi^3L^2}\beta^{2-\frac{2}{\alpha}}\log \left(\frac{\xi}{\sigma}\right).  
\end{align}

The contribution associated with b-diagram  is 
\begin{eqnarray}
I^{(b)}(\alpha,\xi)&=&-\frac{3\ell^2}{\pi^3} \int_{-\beta}^\beta d\vartheta   \int d^2\tilde{\sigma} \int_{-\beta}^\beta d\vartheta' \frac{(\vartheta'- \vartheta)^2- \tilde{\sigma}^2}{\left[(\vartheta'- \vartheta)^2+ \tilde{\sigma}^2\right]^2}g\left[\vartheta^{-1}(\tilde{\sigma})\right] \nonumber \\
&\simeq&\frac{6\ell^2\beta}{\pi^3} \int_{-\beta}^\beta \frac{d\vartheta}{\beta^2 -\vartheta^2}  \left|\frac{\partial \vartheta}{\partial z} \right|^{2}\simeq \frac{6\ell^2\alpha^3\beta^{2-\frac{2}{\alpha}}}{\pi^{3} L^2}\log\left(\frac{\xi}{\sigma}\right)
\end{eqnarray}
where to obtain this result we kept only terms which are proportional to the logarithm. This contribution has the same form and the same sign of the previous one apart from a different prefactor.  

Turning to the calculation of the local contribution, we notice that here one should use the property $g(\bm r)\simeq \ell^{2}/\sigma^2$ for $|\bm r|\ll\sigma$,   which allows us to obtain: 
\begin{equation}
I^{(c)}(\alpha,\xi)= \frac{ \ell^2 \alpha^2\beta^{2}}{\pi^2(2\alpha-1)\xi\sigma}
\end{equation}
\end{subequations}
From Eqs.~(\ref{dd}), (\ref{DeltaAlphaExpression}) and (\ref{Cintegrals}) we obtain that the correction to the fractal dimension, 
\begin{align}
&\delta d=d-d_0=-\frac{1}{2p^2\log(\xi/L)\log(\xi/L_{\max})} \left.\frac{\partial}{\partial \alpha}\left[I^{(a)}(\alpha,\xi)+I^{(b)}(\alpha,\xi)+I^{(c)}(\alpha,\xi)\right]\right|_{\alpha=\alpha^*},
\end{align}
is:
\begin{align}
 &\delta d = \frac{\epsilon^2 \alpha^{*2}}{\log\left(\frac{L_{\max}}{\sigma}\right)-\log x} \left( \frac{42\alpha^{*}  }{\pi}\frac{ \log x}{ x^2} +\frac{1}{ (2\alpha^*-1) }\frac{1}{x}\right), \label{d}
\end{align}
where $x= \xi/\sigma$ is the dimensionless box size. To obtain this result we kept only leading order terms in $\log(L/\xi)$.

Within our perturbative approach we can choose $\alpha^*$ to be the maximal point of $f(\alpha)$ corresponding to the non-disordered system, i.e  $\alpha^*\approx 4$, see e.g. Ref.~\cite{efofalfa}. For this choice it follows that the first term in the right hand side of Eq. (\ref{d}), associated with the non-local contributions, is dominant within the range $1\lesssim x \lesssim 3000$. Outside this regime, the  local contribution (the second term in Eq. (\ref{d}) coming from c-diagram becomes dominant.  However, such a situation seems to be very difficult to realize experimentally. 

\section{conclusion}
  \begin{figure}[t]
~~~~~~~~~~~~~~~~~~~~~~~~~\includegraphics[width=0.6\columnwidth]{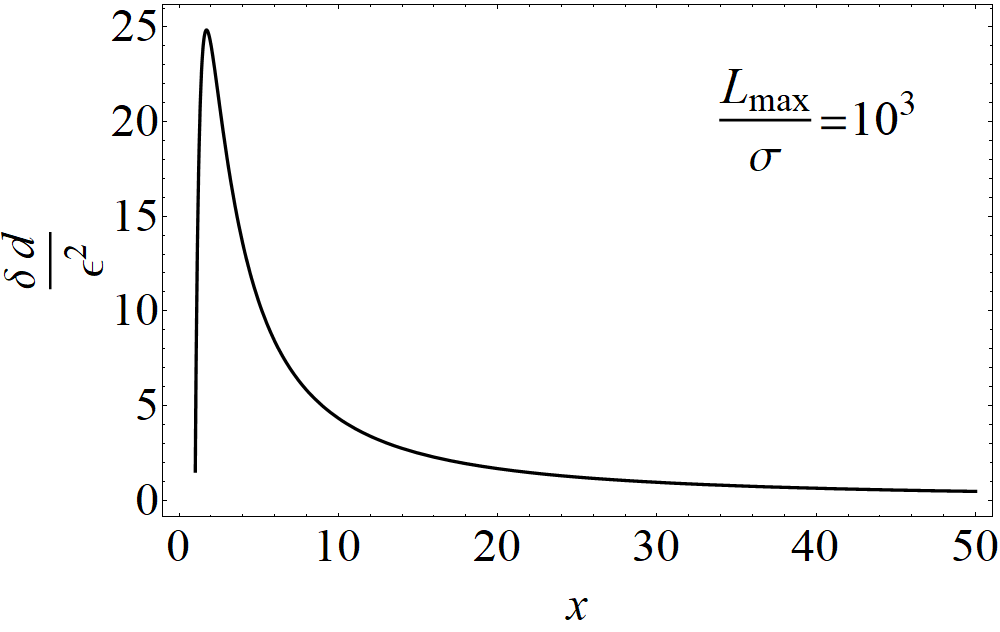}
\caption{The behavior of the correction to the fractal dimension as function of the dimensionless box size $x=\xi/\sigma$.}
 \label{deltadvsxFig}
\end{figure}
To summarize, we have calculated the effect of weak disorder on the fractal properties of the Saffman-Taylor bubble in Hele-Shaw cell. The correction to the fractal dimension, $\delta d$, is expressed in terms of the two-point correlation function of the harmonic moments. These  correlations may be understood within the Kogam-Shul'man approach as generated by a uniform distribution of local point sources \cite{KS}. This picture implies that  disorder drives the system towards the Eden model \cite{Eden}   in which clusters are grown by attachment of new particles randomly along their boundary.     

The behavior of the correction to the fractal dimension as function of the dimensionless box size, $x=\xi/\sigma$ is depicted in Fig.~4. This correction is maximal when the box size becomes close to the disorder correlation length. To the leading order in $\nu=1/\log(L_{\max}/\sigma)$ the maximal value of the correction to the fractal dimension is at $x=x_{\max}\simeq 1.65+ 0.41\nu$, for which $\delta d(x_{\max})\simeq \epsilon^2\nu(159 +  79\nu)$. The correction to the fractal dimension decreases with the box size, up to  a box size of order of the size of  the Saffman-Taylor bubble, $L_{\max}$. For a fixed box size, $\xi$, the fractal dimension converges to the clean system value as the bubble becomes larger but only logarithmically with its size. Namely, disorder in this problem is irrelevant in the RG sense.

\
\
\\

\noindent
{\bf Acknowledgments}

This research was supported by the Israel Science Foundation (ISF) Grants Nos. 302/14 (O.A.) and  1466/15 (E.B.).

\end{document}